\documentclass[authoryear,preprint,review,12pt]{amsart}

\usepackage[centertags]{amsmath}
\usepackage{amssymb}
\usepackage[pdftex]{graphicx} 

\begin{document}


\title{Fixes to the Ryden \& McNeil Ammonia Flux Model}

\author{William M. Briggs}
\address{New York City, NY}
\email{matt@wmbriggs.com}

\author{Jaap Hanekamp}
\address{University College Roosevelt, Middelburg, the Netherlands; Environmental Health Sciences, University of
Massachusetts, Amherst, MA, USA}
\email{j.hanekamp@ucr.nl; hjaap@xs4all.nl}

\begin{abstract}
We propose two simple fixes to the Ryden and McNeil ammonia flux model. These are necessary to prevent estimates from becoming unphysical, which very often happens and which has not yet been noted in the literature. The first fix is to constrain the limits of certain of the model's parameters; without this limit, estimates from the model are seen to produce absurd values. The second is to estimate a point at which additional contributions of atmospheric ammonia are not part of a planned expert but are the result of natural background levels. These two fixes produce results that are everywhere physical. Some experiment types, such as surface broadcast, are not well cast in the  Ryden and McNeil scheme, and lead to over-estimates of atmospheric ammonia. 

\end{abstract}


\maketitle


\section{Introduction}
\label{intro}

Atmospheric ammonia (NH3; we use both interchangeably) concentrations in The Netherlands are reported as amongst the highest in the world and regarded as a threat to biodiversity in nature areas. Livestock is the largest contributor to ammonia emissions. Since 1993, major efforts have been expended to reduce emissions. As a practical approach, the reduction of ammonia volatilization after manure application to farmland, regarded as the largest single emission source, got much attention. In the 1990s, broadcast surface spreading made way for methods such as shallow- and narrow-band injection on grassland and deep placement on arable land (fallow).

However, as the 2015 Sutton-review on the scientific underpinning of calculation of ammonia emission and deposition in the Netherlands remarked, ammonia concentrations in the air ``have not decreased as much as expected since the introduction of mitigation measures. This has led stakeholders to question the effectiveness of the Dutch ammonia policy", (Sutton et al., 2015).

In 2017, we published a critical paper on the ammonia discourse in the Netherlands (Hanekamp et al., 2017), and got responses both from the National Institute for Public Health and the Environment (RIVM henceforth) and the Wageningen University and Research (WUR henceforth), to which we responded subsequently (van Pul et al., 2017; Briggs et al., 2017a; Goedhart and Huijsmans, 2017a, Briggs et al., 2017b; Goedhart and Huijsmans, 2017b). However, what was lost in this flurry of points and counterpoints was the absence of the manuring emission data of the WUR.

Huijsmans and Schils published an overview of experimentally assessed {\it emissions} of ammonia (emission factors) after manure application (Huijsmans and Schils, 2009), referring to 199 experiments done over the years. One of our goals was a reproduction of the presented results using the underlying data. The reasons were on the one hand a straightforward scientific interest and on the other hand a desire to settle a continuous dispute over the published results and the implications for the prevailing agricultural policies in the Netherlands.

It was made clear that all the datasets mentioned in the 2009 paper by Huijsmans and Schils were no longer available. This was quite unfortunate, as the claim of declining ammonia emissions due to regulatory enforced low-emission manure application techniques hinges on these data. Consequently, the results as reiterated by Huijsmans et al. (2015), could not be reproduced. Fortunately, in mid-2018 it was announced that manuring-emission data were still available. We subsequently received these data. The table below describes the categories, types and number of the 160 manure experiments. The manuring experiments span some two decades: from 1997 to 2017. All received experimental data are on file and available.

{\small
\begin{table}[htbp]
  \begin{tabular}{llrr}
Category & Type & Count\\\hline\hline
Surface Broadcast & General & 19 \\
                  & High rate & 8 \\
                  & Low rate & 7\\
Shallow Injection &  General & 62 \\
Narrow Band       &  General & 21 \\
                  & Reference pH & 4 \\
                  & Altered pH & 8 \\
                  & Diluted (normal pH) & 26 \\
Green Duo         & General  & 5 \\\\                  
  \end{tabular}
\caption{The categories and types of the 160 manure experiments.}
\label{tab1}
\end{table}
}

In this paper, we review the model used by the WUR to estimate ammonia fluxes over manured plots, the so called Ryden and McNeil model (1984; henceforth R\&M). Using the manuring data, we will compare NH3-emissions from the different manuring techniques. By doing this comparison, R\&M will be scrutinised. We will show that R\&M fails on multiple levels and should be recast incorporating more accurate boundary layer physics and chemistry. Assuming the structure of the R\&M model will be used, and not entirely recast, a number of model fixes are proposed and will be discussed. Our conclusions imply that policies that prescribe certain manuring techniques as emitting less ammonia than those that need to be discarded are not on sure footing, and express overconfidence. 

\section{Model Fixes}

Ryden and McNeil (1984) offered a quasi-physical model to represent NH$_3$ flux over a manured plot. Recapiulating Ryden and McNeil (1984) and Hanekamp et al. (2017), the ammonia flux equation is
\begin{equation}
\label{RM0}
F = \frac{1}{x} \int_{z_0}^{z_p} \overline{uc}\; dz
\end{equation}
\noindent where $x$ is the fetch of the plot, $z_0$ and $z_p$ the height limits, $u$ the instantaneous wind speed, $c$ the instantaneous value NH$_3$,  and $\overline{uc}$ the time-averaged flux at height $z$.
The solution to (\ref{RM0}) requires knowing the functional relationship between $\bar{uc}$ and $z$. Instead of a physical argument, RM first assumed $u$ and $c$ were independent; they next created two empirical functional relationships between $\bar{u}$ and $\bar{c}$ and height $z$.  The two equations are:
\begin{subequations}
\begin{align}
  \bar{u}(z)  =& D \ln z + E   \label{RM1a},\\
  \bar{c}(z)  =& -A \ln z + B \label{RM1b}.
\end{align}
\end{subequations}

Equations \ref{RM1a} and \ref{RM1b} are substituted into (\ref{RM0}) for $\bar{uc}$ and the integral is then solved. The coefficients $A$, $B$, $D$, and $E$ are unknown but estimated by ordinary linear regression between the natural log of height and the observed wind speed and observed NH$_3$. Equations (\ref{RM1a}) and (\ref{RM1b}) are input into (\ref{RM0}), which is integrated, the end result of which is:
\begin{eqnarray}
\label{RM2}
F &=& \frac{1}{x}\biggr[  
-AD\left(z(\ln z)^2 - 2z\ln z + 2z\right) + (BD-AE)\left(z(\ln z -1) \right) \nonumber \\
 && + EBz-\bar{c}_1D\left(z(\ln z - 1)\right) - \bar{c}_1Ex\biggr]\biggr|_{z_0}^{z_p},
\end{eqnarray}
\noindent where $\bar{c}_1$ is the ambient or background concentration of NH$_3$ on the windward side of the site, and $z_0$ and $z_p$ are the difference in heights. 

Crucially, Ryden and McNeil solve for $z_0$ and $z_p$ by setting $\bar{u}(z)=0$ in (\ref{RM1a}) and setting $\bar{c}(z)$ equal to the windward side background NH$_3$ concentration $\bar{c}_1$. Thus 
\begin{subequations}
\begin{align}
  z_0  =& \exp{(-E/D)}   \label{RM1a2}\\
  z_p  =& \exp{((B-\bar{c}_1)/A)} \label{RM1b2}
\end{align}
\end{subequations}

The empirical relationship (\ref{RM1a}) is known to hold well (see Stull, 1988); estimates of $z_0$ are robust across all datasets we checked. But as vertical concentrations of NH$_3$ approach the windward background concentration, the relationship (\ref{RM1a}) begins to fail, as will be demonstrated below. As a result, estimates of $A$ can be at or near 0, which causes $z_p$ to explode.  This in turn, as is clear from (\ref{RM2}), causes estimates of $F$ to also explode.  

\begin{figure}[htbp]
        \centering
        \includegraphics[scale=.5]{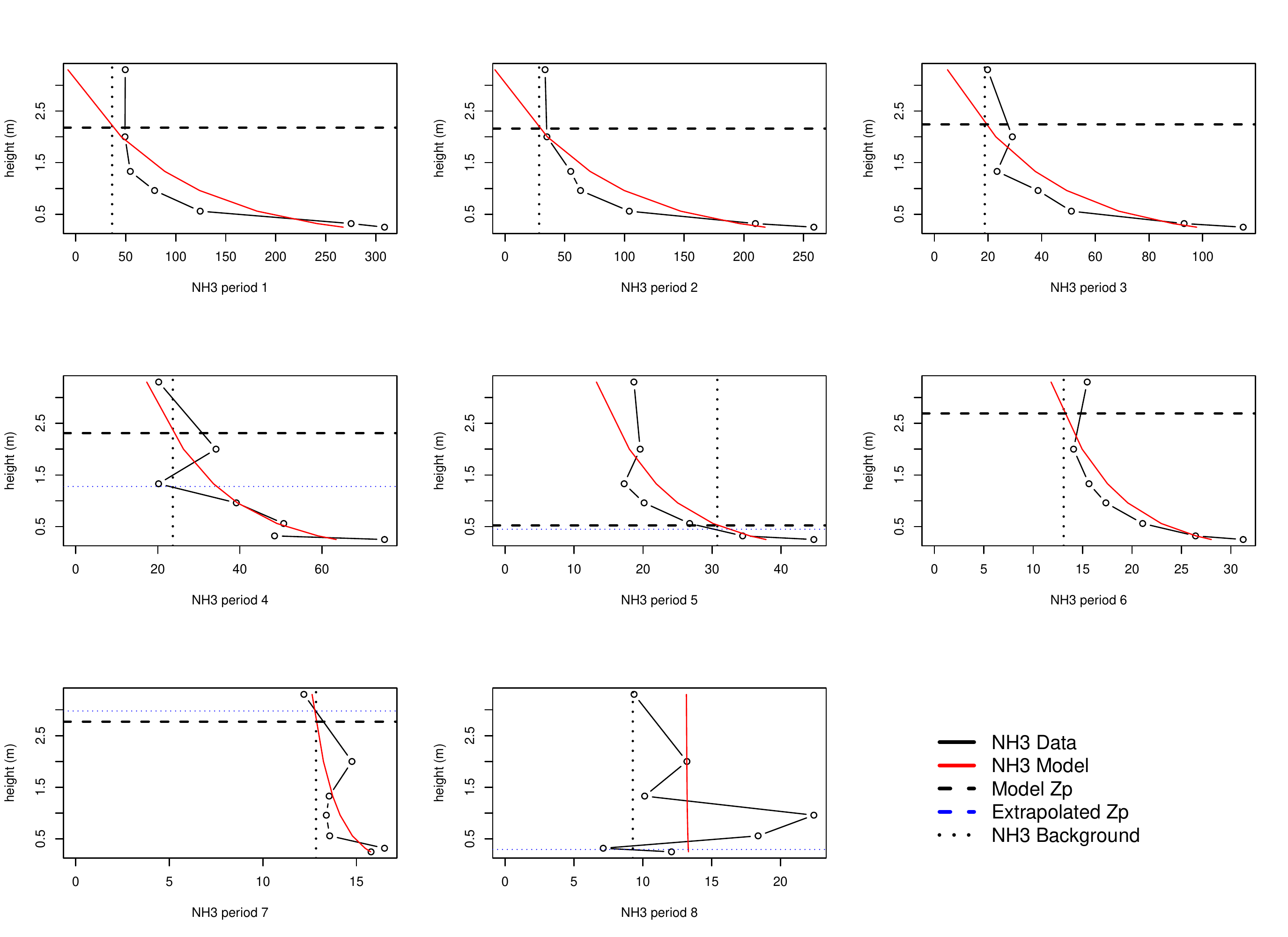}
\caption{\label{fig1} The values of NH$_3$ for several periods for MarkeDuiven 1999 (W19) Field 2. The vertical axis is height (m); the horizontal is measured  NH$_3$. The black lines with open circles are the measured values of  NH$_3$. The red line is the regression equation (\ref{RM1b}). The vertical black dotted line is the measured windward background concentration of NH$_3$. When seen, the horizontal black dashed line is the calculated value of $z_p$ by (\ref{RM1b2}). When seen, the horizontal dashed blue line represents a more robust estimate of $z_p$ as explained in the text.}
\end{figure}

Here is an illustration how this happens. Experimental data is taken from MarkeDuiven 1999 (W19), field 2, which ran a shallow injection experiment. Data was collected over eight periods, from 1 May 1999 8:20 AM until 14 May 1999 10:40 AM. Each collection period lasted about one to two hours. A profile of the data collected and the modeled values appears in Fig. \ref{fig1}.

The vertical axis is height (m); the horizontal is NH$_3$ concentrations (in $\mu$g/m$^3$) . The black lines with open circles are the measured values of  NH$_3$. The red line is the regression equation (\ref{RM1b}) fit to the data. The vertical black dotted line is the measured windward background concentration of NH$_3$. When seen, the horizontal black dashed line is the calculated value of $z_p$ by (\ref{RM1b2}). When seen, the horizontal dashed blue line represents a more robust estimate of $z_p$ to be described below.

Immediately after the manure is applied, in Period 1, the model provides a ballpark job of following the data, though it over-estimates NH$_3$ concentrations over most of the height range. This analysis is before we consider the uncertainty of the model's regression coefficients, which we do below. The model assumes a more gradual reduction in NH$_3$ concentrations by height than is observed, a typical outcome. Nevertheless, the estimate $z_p$ does comes at a height where the  NH$_3$ concentrations appears to be in equilibrium with its surroundings, and about equal to the background concentration. 

The same is true for Period 2. But by Period 3, the model estimates $z_p$ at a point higher than where the measured data is close to the background concentration. The difference is a little less than 1 meter. This is a large difference considering the maximum mast height is placed at just over 3 m.  Period 4 shows a similar discrepancy. The model is more or less good until Period 8, when it breaks down completely.  The estimate for $A$ is near 0, hence the regression model line is almost vertical. The estimate for $z_p$ is thus near machine computational limits, which causes the esimate of $F$ ot blow up. The reason is obvious. The observed data is already at or near the background concentration, with only natural variation by height. Meaning the regression approximation breaks down. This insight leads to two possible solutions.

The first proposed fix is provided in the dashed blue line. This represents a simple linear extrapolation of where the measured values of NH$_3$ intersect the observed background concentration (where the curves cross), when and if this happens; it sometimes does not. The idea is that once the measured leeward NH$_3$ is equal to the windward background concentration, values above this height likely represent natural variation.  In Period 8, the observed NH$_3$ crosses the background concentration at less than 0.5 m, a height which provides a far better estimate of $z_p$ than the regression model's estimate (which is absurd).  

Thus we select in each period the esimate of $z_p$ that is lower, the R\&M regression estimated height or the extrapolated intersection.  This better represents how the actual data behaves than the regression model alone, which is unstable.  

The second proposed fix involves incoporating the uncertainty in the regression model. Now it is clear Period 8, even assuming a reasonable estimate of $z_p$, should not be used in estimating total NH$_3$ emissions from a manuring. This is because the data is already at or near background level with some natural variability.  It would be a mistake to continue adding to the total emissions from a manuring those values which are genuine background. Doing so inflates estimates.  This is concluded from examining Period 7, which already had lower NH$_3$ concentrations across its range, values which had been steadily decreasing since Period 1. The extrapolated background-crossing estimate of $z_p$ (blue line) is also near the surface. 

There is uncertainty in the regression coefficients, $A, B, D, E$, which when accounted for produces a range in estimates of $F$.   Methods to account for the uncertainty in the regression coefficients and in $F$ have been given in Goedhart and Huijsmans (2017) and Hanekamp et al. (2017). We use the method of Goedhart and Huijsmans to avoid any dispute over the estimates in coefficients (the Goedhart and Huijsmans method generally provides narrower bounds than Hanekamp; see Briggs et al. 2017a and 2017b).  The Goedhart and Huijsmans method is a bootstrap-like simulation to produce replications of $A, B, D, E$ according to the uncertainty given in the regression model. These parameter samples in turn cause variations in estimates of $z_p$ and $z_0$, and, finally, $F$.   

Another way to think of it is that when the point estimate of the regression coefficient $A$ is close to 0, there is a chance individual replications of $A$ will be less than 0. In a Bayesian sense, the posterior probability of $A<0$ is not small.  When any $A$ is in fact $<0$, not only will estimates of $z_p$ be suspect, but emissions $F$ in (\ref{RM2}) can be negative, which is of course impossible in practice.

We thus to re-calculate the regression (\ref{RM1b}) in a Bayesian context, using ``refererence" priors, and to use this model to calculate the posterior probability of $A<0$. Here, we use the MCMCregression function from the R package MCMCpack, using default (normal) priors.  When this modeled posterior probability of $A<0$ exceeds some threshold, it is determined this period's contribution to emissions, and all future periods' contributions, are essentially background and should not be used. There is no a priori argument for an optimal probability threshold. One that avoids $F<0$ is certainly warranted. We found a threshold probability of 2/3 to work well in practice. None of the results below are essentially sensitive to this number.   

\section{Results}

As stated, the data we received comprised of 160 experiments within three categories and several types from 1997 to 2017, as illustrated in Table \ref{tab1} above.

These experiments ran over six to nine periods of data collection which took place over the course of about one to four days. Central mast heights ran from about 0.2 to about 3.4 m, with measurements taken over five to seven roughly even height increments. Plot fetches, from windward to leeward sides, were about 20 m. 

Emissions of NH$_3$ per period are calculated by subtracting the flux at $z_p$ from the flux at $z_0$, divided by the fetch, with all multiplied by the length of time of the period.  Total emissions per experiment are found by summing emissions across all periods.  Since it is the late periods that often result in estimates of $A$ near 0, and hence periods $F<0$ or very large $F$, it is the late periods that cause enormous (positive or negative) estimates of total emissions.    

The other calculation of interest is the NH$_3$ emitted as a fraction of that applied in the experiment. This is found by dividing the amount applied by the surface area in square meters. Both of these numbers are supplied in the data (with no uncertainty indicated). This quantity is divided into the total emission to provide the percent of NH$_3$ emitted. 

\begin{figure}[htbp]
        \centering
        \includegraphics[scale=.5]{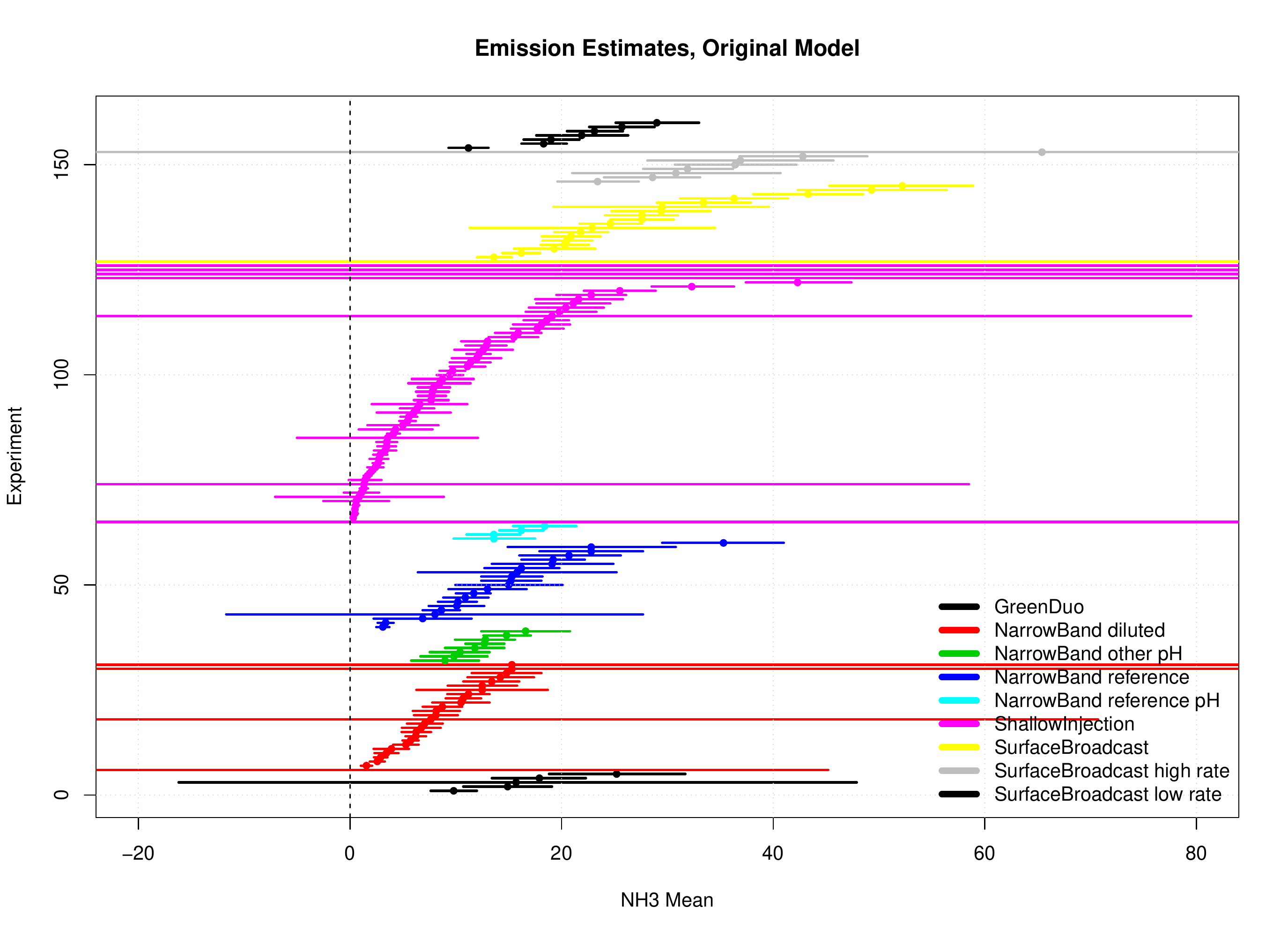}
\caption{\label{fig2} The central estimates of NH$_3$ in kg/hectare and 95\% confidence intervals for each experiment category and type.  Each type has been sorted from low to high by its central estimate, as a guide to the distribution of values. A vertical line at 0 for NH$_3$ has been drawn. It is seen several estimates blow up, caused by near 0 estimates of $A$.}
\end{figure}

Fig. \ref{fig2} shows the results of applying the flux model to each experiment, separated by cateogory and type (Green Duo are the bottom-most, and Surface Broadcast the top-most lines). In each experiment, a dot represents the central estimate of NH$_3$ total emissions, with its 95\% confidence interval shown with a horizontal line, calculated by the method of Goedhart and Huijsmans (2017). 

The figure shows the central estimates of NH$_3$ in kg/hectare and 95\% confidence intervals for each experiment category and type. The experiment number is only a reference to the figure.  Each type has been sorted from low to high by its central estimate, which provides a guide to the distribution of values. A vertical line at 0 for NH$_3$ has been drawn; values less than 0 are obviously impossible. It is seen several estimates blow up, caused by near 0 estimates of $A$. Several of the lower bounds of the confidence intervals are also below 0, even if the central estimates appear reasonable.  The blow ups appear across all categories. 

\begin{figure}[htbp]
        \centering
        \includegraphics[page=2,scale=.5]{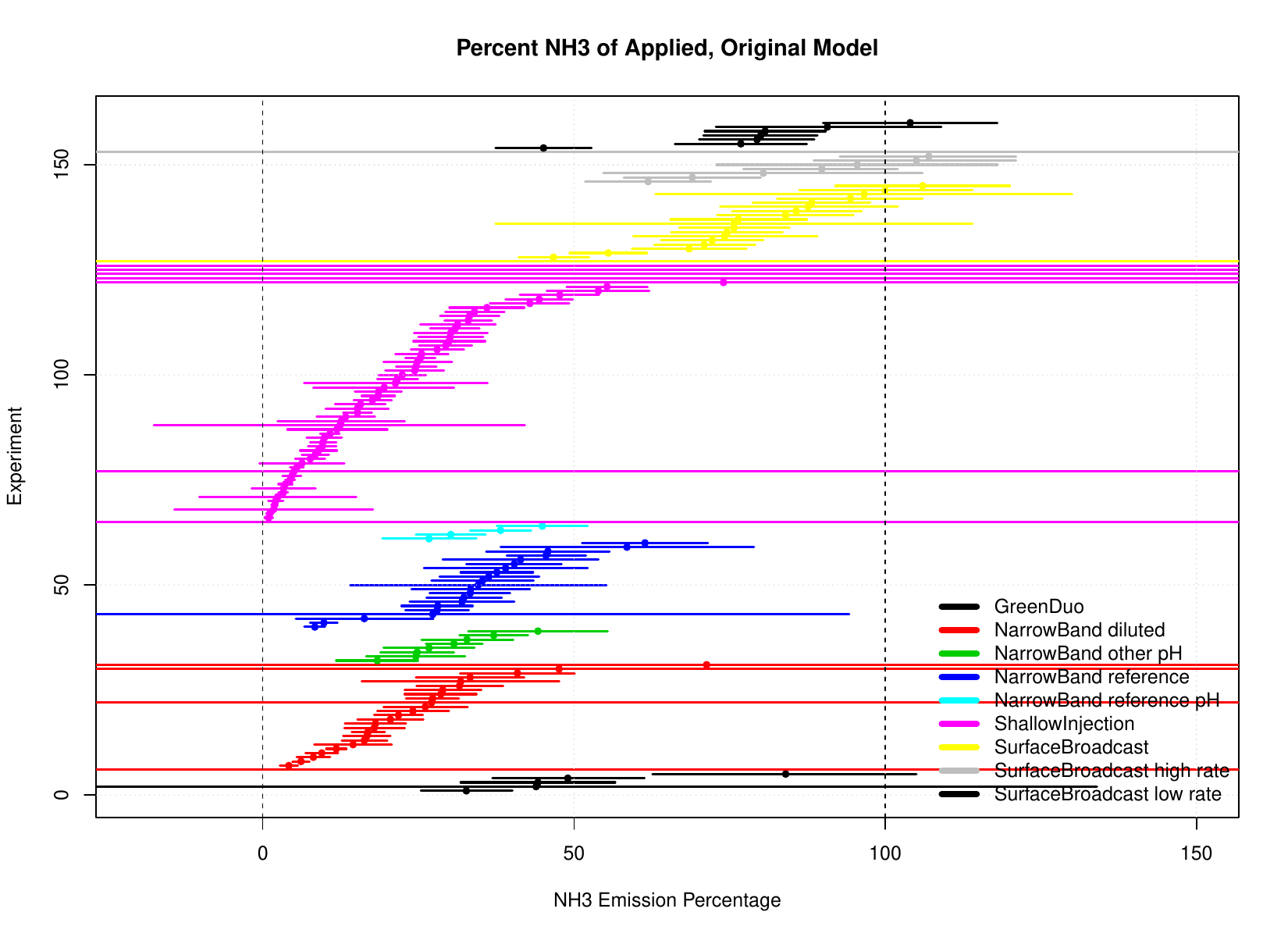}
\caption{\label{fig3} The total emission percentages of experimentaly applied NH$_3$ and 95\% confidence intervals for each experiment category and type.  Each type has been sorted from low to high by its central estimate, as a guide to the distribution of values. A vertical line at 0 and 100\% has been drawn. Some of the non-physical results are caused by near 0 estimates of $A$; others are caused for other reasons explained in the text.}
\end{figure}

There are additional difficulties with the flux model, as demonstrated in Fig. \ref{fig3}. This is the same as Fig. \ref{fig2}, except for emission percentages of experimentally applied NH$_3$. A second vertical line of 100\% has been drawn. It is, of course, not impossible for more NH$_3$ to be emitted than was applied, but only if this additional NH$_3$ came from sources not controlled in the experiment. There is reason, however, to suspect this is not the case here. We diagnose other possible reasons below after showing the effects of the flux model modifications metioned in Section 2. The blow ups and too-wide confidence intervals are seen here as they were in Fig. \ref{fig2}.

We apply the proposed modifications next. 

\begin{figure}[htbp]
        \centering
        \includegraphics[page=1,scale=.5]{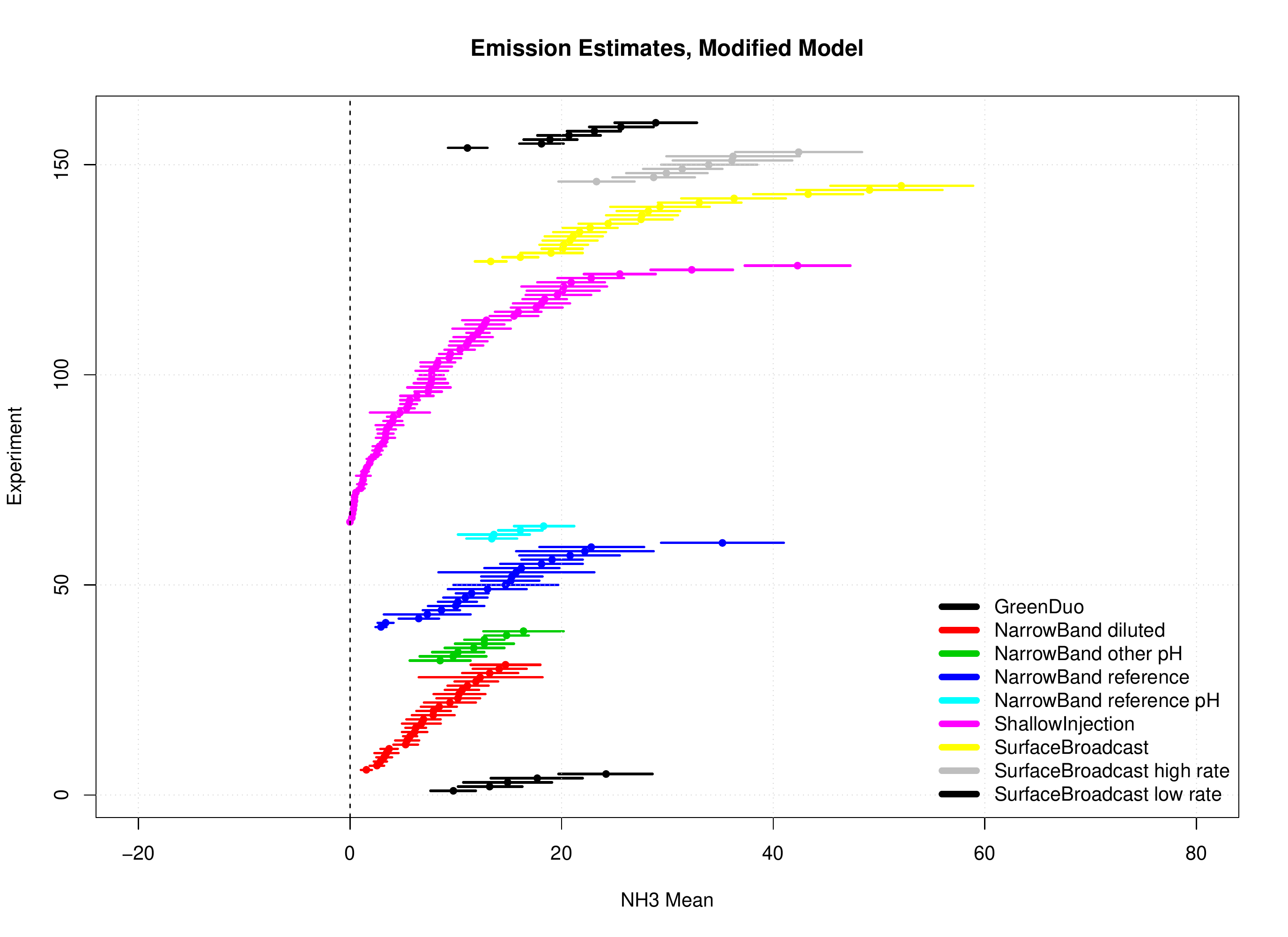}
\caption{\label{fig4} The modified flux model NH$_3$ emissions, after applying both fixes. The notations are identical to Fig. \ref{fig1}.}
\end{figure}

Figure 4 shows the emission total estimates using the modified flux model. The notations are identical to Fig. \ref{fig1}.  The scaling of the figure is also unchanged for ease of comparison. All values are physical for all experiment types. No central flux estimate for any type was less than 0. 

A comparison of the original and modified flux model is in order.  The original unfixed R\&M model had 7 of 160 or 4.4\% of experiments with estimated total emissions less than 0 (5 experiments) or greater than 100 $\mu$g/m$^3$ (2 experiments, with the totals being at least 1000 $\mu$g/m$^3$). The modified model, applying both proposed fixes, has no total estimates less than 0, and no experiments with totals greater than about 50 $\mu$g/m$^3$. The original model had 20 confidence intervals for total emissions less than 0 or greater than 100 $\mu$g/m$^3$. The modified model had none of these. This is summarized in Table \ref{tab2}.

{\small
\begin{table}[htbp]
  \begin{tabular}{llrr}
Unphysical Results & Original model & Modified Model \\\hline\hline
Point estimates & 7/160 (4.4\%) & 0/160 (0\%) \\
Confidence intervals & 20/160 (12.5\%) & 0/160 (0\%) \\
  \end{tabular}
\caption{Comparisons of the original, unmodified and the modified R\&M models of unphysical point estimate results.}
\label{tab2}
\end{table}
}

Excluding the 7 unphysical values in the original model, and comparing the remaining central estimates of emission totals, shows the original model gives estimates of totals about 4.6\% higher on average. This over-estimation comes in part by those times in the original model where essentially background emissions are added to the totals, as explained above. But the largest contribution to the change is in the more physical estimates of $z_p$.  We saw that 119/153 or 77.7\% of the modified model results were less than the original model; 26/153 of 17\% of cases saw matching estimates for both models; only 8/153 or 5.2\% of estimates were smaller in the original model than the modified, with the largest differences in the smallest totals; e.g. one original model estimated a total of 0.86 and the modified model estimates nearly double this at 1.4. There was no pattern in the type of experiment.   Overall, the modified model produced smaller estimated totals on average.  This, and the results about confidence intervals next, are summarized in Table \ref{tab3}.

The next comparisons are in the lengths of the confidence intervals. The original model had wider confidence intervals in 154/160 or 96\% of experiments. In the 6 times the modified model had wider confidence intervals, the difference was about 1\%.   There were 18 cases where the interval was 10 times wider in the original model. Overall, the median reduction of confidence interval width was about 6\%. The mean was not used as a comparison because some of the confidence intervals in the original model were infinite (or at machine limits). 

{\small
\begin{table}[htbp]
  \begin{tabular}{llr}
\multicolumn{2}{c}{Point estimates}\\\hline\hline
Overall & Modified model 4.6\% lower mean point estimates \\
Details & Original $>$ Modified  119/153 (77.7\%) \\
        & Original $=$ Modified  26/153 (17\%) \\
        & Original $<$ Modified  8/153 (5.2\%) \\
\multicolumn{2}{c}{Confidence intervals}\\\hline\hline
Overall & Modified model 6\% median narrower intervals \\
Details & Original $>$ Modified  154/160 (96\%) \\
        & Original $=$ Modified  0/160 (0\%) \\
        & Original $<$ Modified  6/160 (4\%) \\
  \end{tabular}
\caption{Comparisons of the original, unmodified and the modified R\&M models of flux estimates, excluding the 7 unphysical original model results for the point estimates.}
\label{tab3}
\end{table}
}

\begin{figure}[htbp]
        \centering
        \includegraphics[page=2,scale=.5]{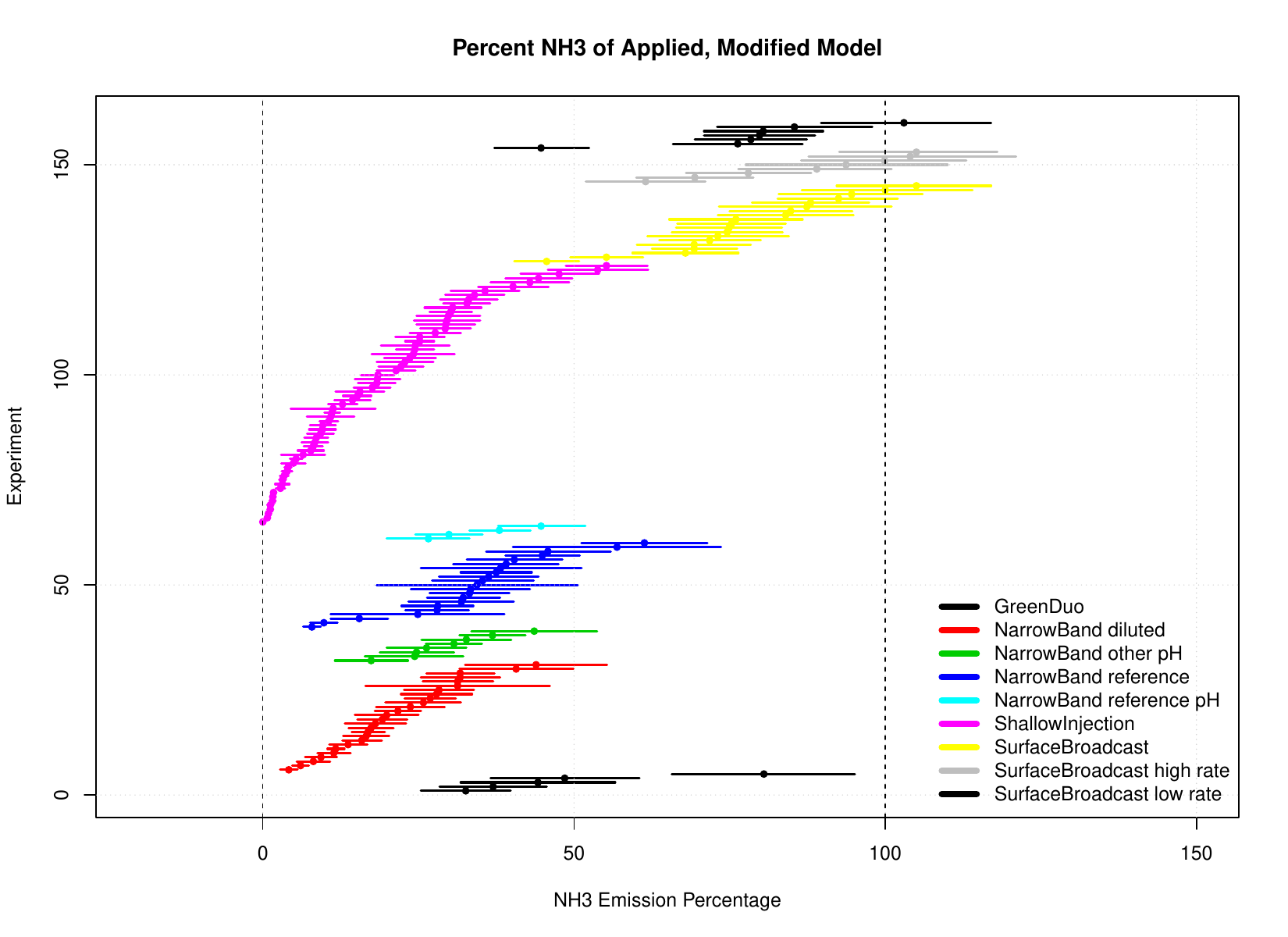}
\caption{\label{fig5} The modified flux model NH$_3$ percent emissions, after applying both fixes. The notations are identical to Fig. \ref{fig2}.}
\end{figure}

Next we examined the modified model's estimates of the percent emissions of NH$_3$ in Fig. \ref{fig5}.

As with the emissions, the modified model largely fixed the problems of blow ups and impossible values. There are still some cases where emission percentages are greater than 100\% but only with the Surface Broadcast cateogory, a point we address below. 

There were 3/160 or 1.9\% of cases where the original model estimated negative percentages. Unphysical results are summarized in Table \ref{tab4}. Excluding these, the estimates were about 7\% lower in the modified model overall. Only 6/150 or 3.8\% of the modified model gave higher percentages; as before, these were for the smallest emission totals. Only 8/150 or 5.3\% of the modified models had longer confidence intervals; the rest of 95.2\% were wider in the original model. The median was about 6.6\% wider; as before, the mean is a bad estimate because of the blow ups.  Again, the modified model produces lower and more certain estimates than does the original model. These results are summarized in Table \ref{tab5}.

{\small
\begin{table}[htbp]
  \begin{tabular}{llrr}
Unphysical Results & Original model & Modified Model \\\hline\hline
Point estimates & (3+9)/160 (7.5\%) & (0+4)/160 (2.5\%) \\
Confidence intervals & (20+28)/160 (17.5\%) & (0+11)/160 (6.9\%) \\
  \end{tabular}
\caption{Comparisons of the original, unmodified and the modified R\&M models of unphysical percentage manure results. The numbers are $(x+y)/160$, where $x$ are the number $<0$ and $y$ the number $>100$. In the modified model, all unphysical results were from Surface Broadcast experiments.}
\label{tab4}
\end{table}
}

{\small
\begin{table}[htbp]
  \begin{tabular}{llr}
\multicolumn{2}{c}{Point estimates}\\\hline\hline
Overall & Modified model 7\% lower mean point estimates \\
Details & Original $>$ Modified  154/160 (96.3\%) \\
        & Original $=$ Modified  0/160 (0\%) \\
        & Original $<$ Modified  6/153 (3.7\%) \\
\multicolumn{2}{c}{Confidence intervals}\\\hline\hline
Overall & Modified model 6\% median narrower intervals \\
Details & Original $>$ Modified  152/160 (95\%) \\
        & Original $=$ Modified  0/160 (0\%) \\
        & Original $<$ Modified 8/160 (5\%) \\
  \end{tabular}
\caption{Comparisons of the original, unmodified and the modified R\&M models of flux percentage estimates.}
\label{tab5}
\end{table}
}

The question remains why some estimates of emission percentages are larger than 100\% in the modified model, and this only found in Surface Broadcast experiments. To answer this we examine more closely the measured windward background concentrations as a function of time. Ideally, these would all be at the ambient background level, with variation provided by natural sources independent of the manure experiment. This is not so in practice, however. 

\begin{figure}[htbp]
        \centering
        \includegraphics[page=1,scale=.5]{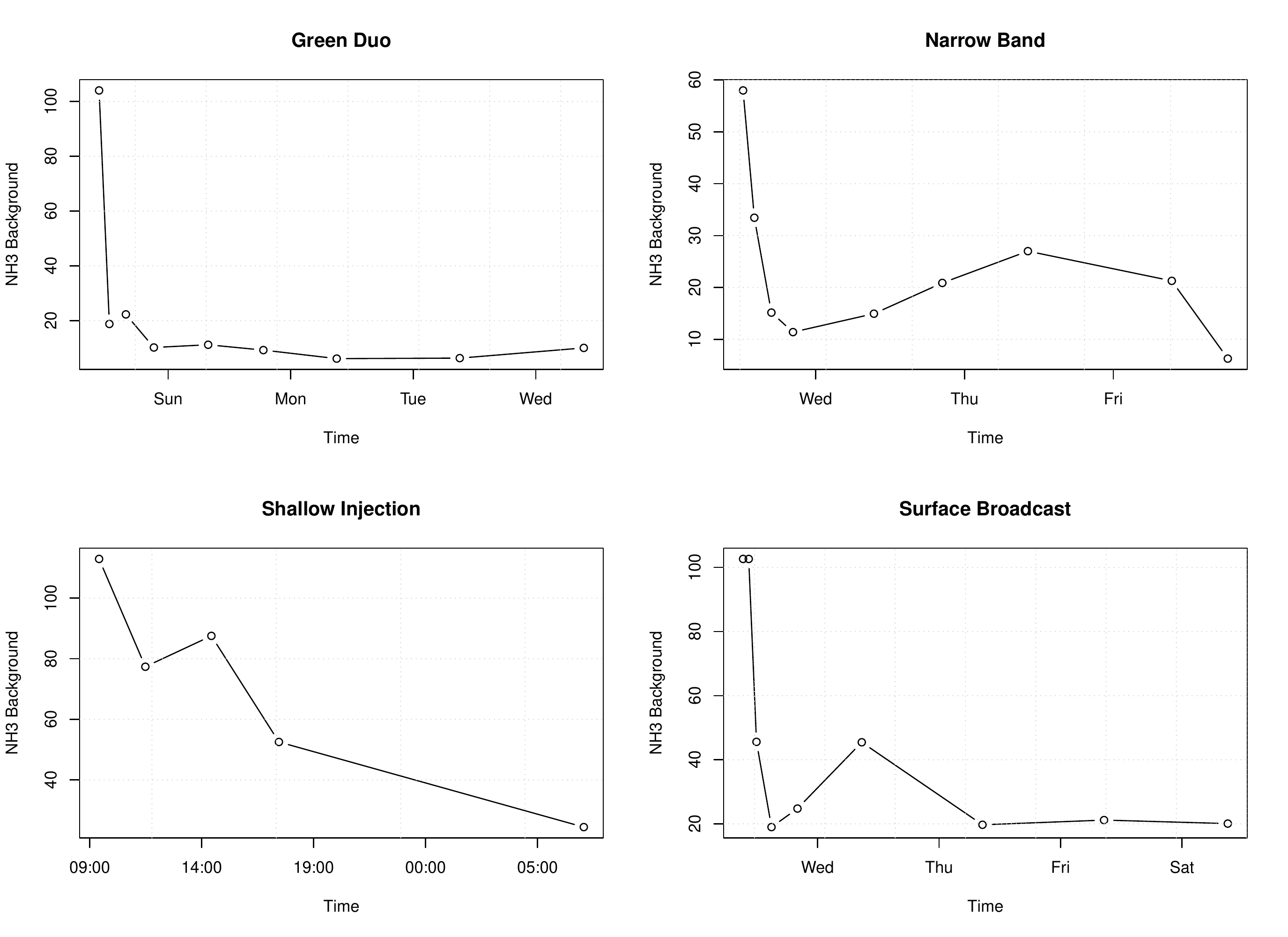}
\caption{\label{fig6} The time series of windward side background conncentrations for four typical experiments, one of each category. Each begins high and more or less rapidly descends to ambient background.}
\end{figure}

\begin{figure}[htbp]
        \centering
        \includegraphics[page=1,scale=.5]{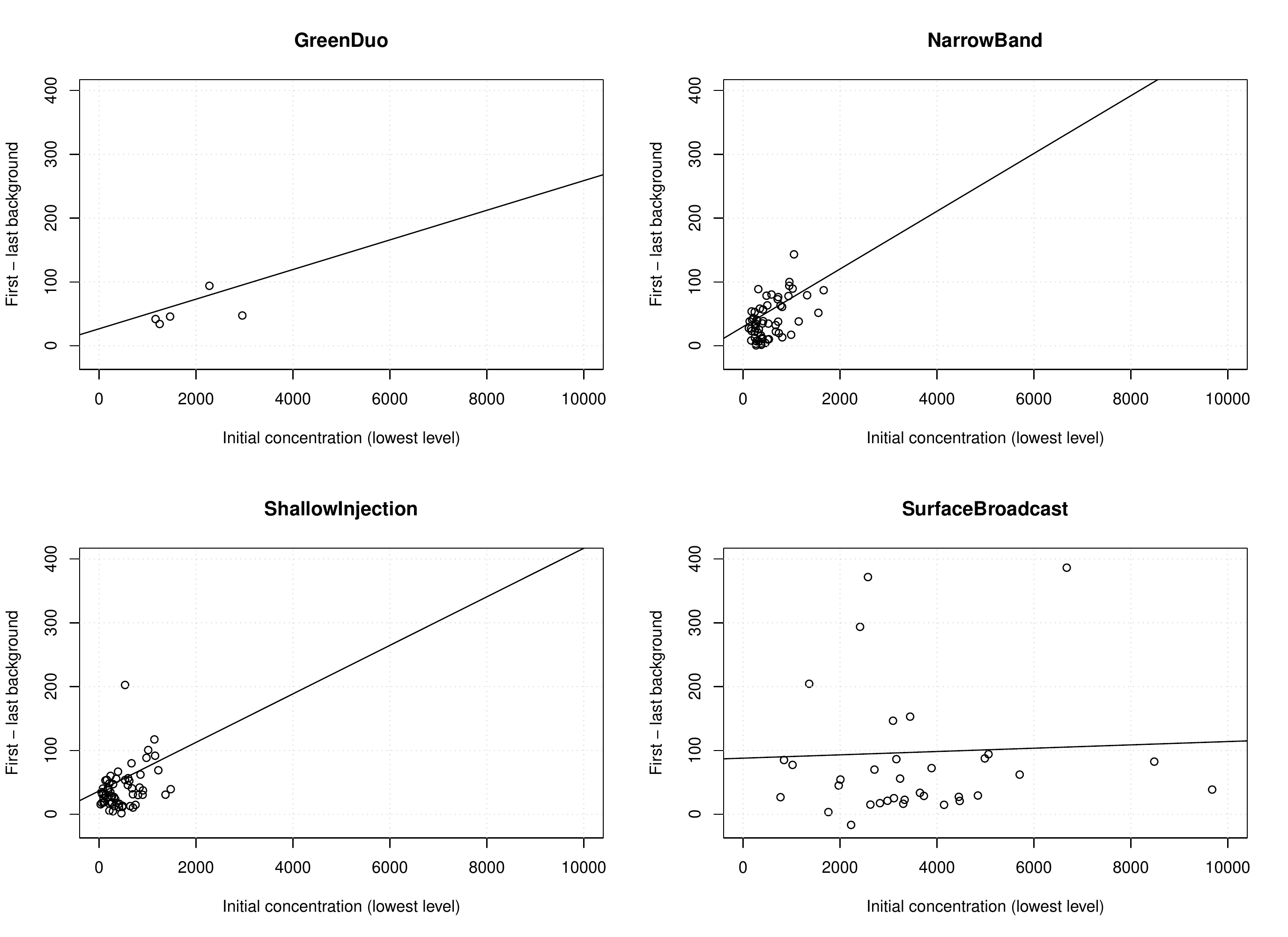}
\caption{\label{fig7} The initial concentration of NH$_3$ at the lowest level by the first minus last windward side background concentration. A simple linear regression has been added in each case.}
\end{figure}

Ideally the windward side background concentration should be at ambient levels, with variation provided by local environmental circumstances. The experiment itself should not cause the windward background levels to increase. It does, however. Evidence is provided in Figs. \ref{fig6} and \ref{fig7}.

Fig. \ref{fig6} shows the time series of windward side background conncentrations for four typical experiments, one of each category. Each begins high and more or less rapidly descends to ambient background. In turn, the data were from the Green Duo 2016 week 27 Green Duo Proef 6, Narrow Band 2016 week 20 Verdunnen Proef 3, Shallow Injection 2000 MarkeDuiven W19, and Surface Broadcast 1997 DGI W33. The time of the start to finish Shallow Injection experiment is not a mistake; this short period is the entire run of the experiment.

Fig. \ref{fig6} shows that as the amount of nitrogen added by the experiment, as measured by the proxy of the first NH$_3$ reading at the lowest height, the change in the first to last windward background concentration increases for all categories, though the effect is different for Surface Broadcast. The types did not matter in these plots, so all data was collapsed to its parent category. If the experiment did not effect the background levels, we would expect the first minus last values to be centered at 0, with roughly equal departures on either side of 0 representing natural variability.  Clearly, there is some ``leakage" or spillage of NH$_3$ toward the windward dectectors. Since wind direction tends to turn with height in the boundary layer and turbulence is present (Stull 1988), it is not surprising some NH$_3$ from the experiment would be blown toward the (what is labeled in the experiment as the) windward side.  The average decrease in background concentration is 46 $\mu$g/m$^3$ for Green Duo, 35 for Narrow Band, 31 for Shallow Injection, and 50 for Surface Broadcast.

The flux model, however, does not account for this drop.  This is not a large problem for the flux estimates themselves, though. This is because the flux measures the amount passing through a certain area relative to the background where the additions, and therefore total emissions, come from the plot and not the background. Emissions are relative, too. 

But there is something else different about Surface Broadcast methods. It is not in the amount of NH$_3$ estimated to have been added in the manure experiments, which did not differ much on average by category. The mean kg added were 5 for GreenDuo (minimum 4.1, maximum 6), 5.8 for Narrow Band (minimum 2.8, maximum 10.4), 7.2 for Shallow Injection (minimum 3.2, maximum 15.3), and 6.2 for Surface Broadcast (minimum 3, maximum 11.3). Yet initial concentrations were much higher for Surface Broadcast: a mean of 1822 for Green Duo, 531 for Narrow Band, 467 for Shallow Injection, and 3550 for Surface Broadcast, i.e. roughly 10 times higher on average.  

\begin{figure}[htbp]
        \centering
        \includegraphics[page=1,scale=.5]{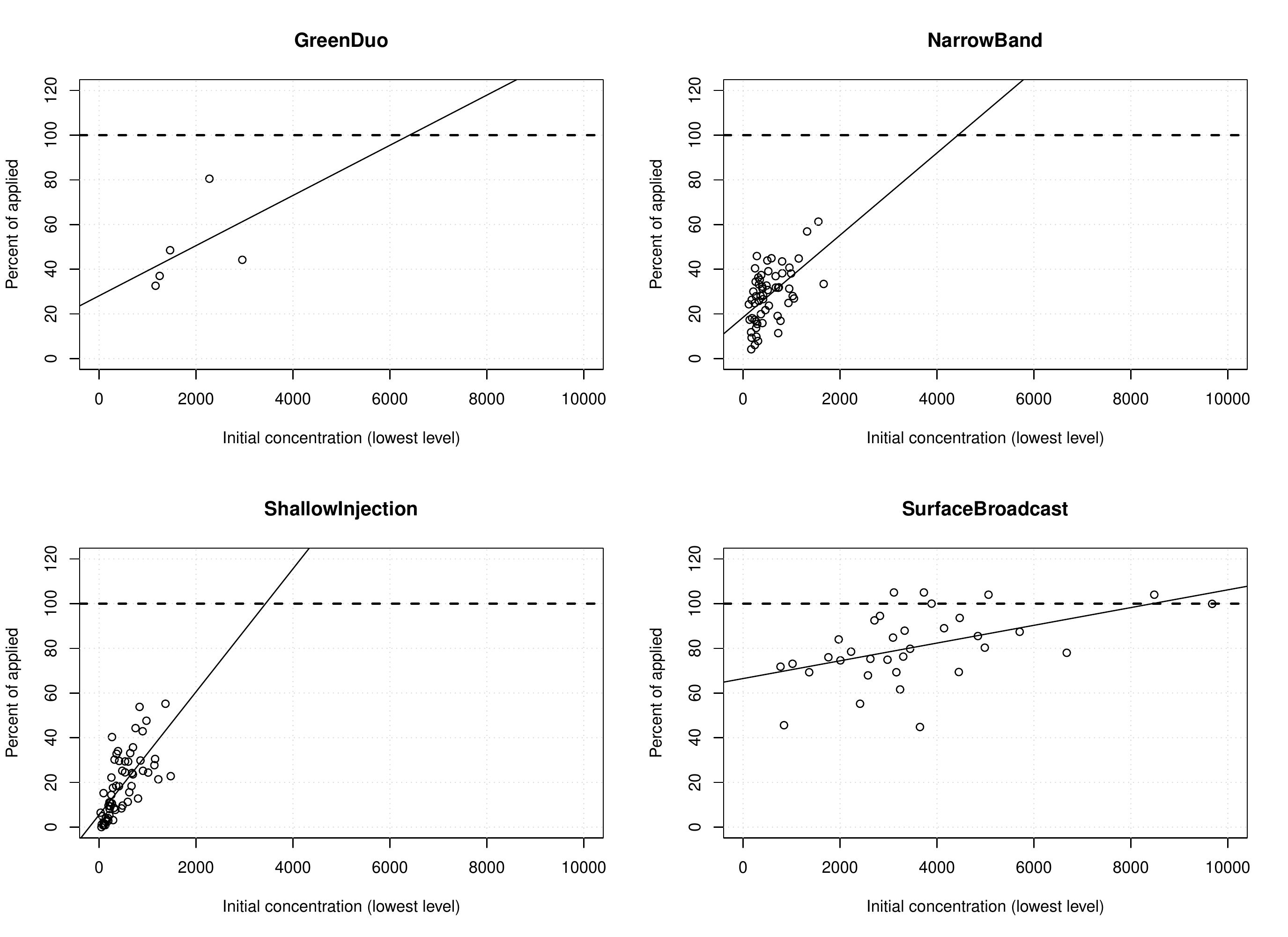}
\caption{\label{fig8} The initial concentration of NH$_3$ at the lowest level by percent applied. A simple linear regression has been added in each case.}
\end{figure}

There is another difference, which can be seen in the bottom right panel of Fig. \ref{fig6}. There is a data artifact in Surface Broadcast station observations: the first value of the background is repeated (exactly) in the second value for all experiments (see the two points by 100 m on the graph for an example). That is, there is a duplicate record, or what appears to be a duplicate record, at all stations, differing only in the time stamp. It would be expected the second value would be smaller than the first. Since it is not, it leads to an inflation in background rates. However, the effect this has on flux is somewhat minimal since flux is relative.  Fig. \ref{fig8} better highlights the difficulty with Surface Broadcast experiments. At the least, this duplication is more evidence higher quality experiments are needed for assessing the true effect of Surface Broadcast.

The figure shows the eventual estimated percent applied as a function of the initial measured NH$_3$ at the lowest level. A similar plot of percent by amount of NH$_3$ applied does not show as strong a relationship.  It is notable Surface Broadcast experiments evince a much higher on average first measured concentration, doubtless due to the manure's direct exposure to the air. These large initial measurements boost the emission totals, and combined with the other model approximations, produce the over-estimates.  The problem can be alleviated in part by more careful measurements, and in part by investigating a more complex physical flux model. We do not attempt the latter here, since there would be no data to evaluate such a model.

\section{Conclusion}

We have shown that the R\&M model contains numerous flaws. We proposed two fixes which solve a number of difficulties of the R\&M model; these difficulties are the ill-fit of the NH$_3$ and wind by height statistical relationship which causes unphysical results, and the tendency to add NH$_3$ background rates to emission totals.  

The regression model (\ref{RM1b}) derived in the R\&M cannot be said to fit well in practice, even with the fixes applied. Obviously, some empirical relationship is necessary if (\ref{RM0}) is to have an analytic solution. It would be best in future work if the R\&M model itself was recast to incorporate more accurate boundary layer physics and chemistry. 

Apart from the modelling itself, the experimental setups as found especially in the surface broadcast experiments show serious flaws. The surface broadcast experiments were done poorly: even the data collection needs better accounting.  There is also the leakage of ammonia found in the background concentrations, which started much too high to be considered true background. The effect on the calculated emissions of this technique are relative, as flux is relative, yet because of these difficulties, they are not easily estimated precisely. The flaws in the experiments ares directly related to the R\&M model, which is too crude to estimate the overall effect of the quality of the  experiments themselves. 

This is serious, as the results generated by the R\&M model, including the fixes, do not give us outcomes that are reliable expressions of actual NH$_3$ emissions.



\end{document}